# Towards Evidence-Based Ontology for Supporting Systematic Literature Review

Yueming Sun[1,2], Ye Yang[1], He Zhang[3], Wen Zhang[1], Qing Wang[1]
[1]The Lab for Internet Software Technologies
Institute of Software, Chinese Academy of Sciences, Beijing 100080, China
[2]Graduate University of Chinese Academy of Sciences, Beijing 100039, China
[3]National ICT Australia, University of New South Wales, Sydney, Australia
{sunyueming, ye, zhangwen, wq}@itechs.iscas.ac.cn, He.Zhang@nicta.com.au

*Abstract*—**[Background]: Systematic Literature Review (SLR) has become an important software engineering research method but costs tremendous efforts. [Aim]: This paper proposes an approach to leverage on empirically evolved ontology to support automating key SLR activities. [Method]: First, we propose an ontology, SLRONT, built on SLR experiences and best practices as a groundwork to capture common terminologies and their relationships during SLR processes; second, we present an extended version of SLRONT, the COSONT and instantiate it with the knowledge and concepts extracted from structured abstracts. Case studies illustrate the details of applying it for supporting SLR steps. [Results]: Results show that through using COSONT, we acquire the same conclusion compared with sheer manual works, but the efforts involved is significantly reduced. [Conclusions]: The approach of using ontology could effectively and efficiently support the conducting of systematic literature review.**

## I. INTRODUCTION

Systematic Literature Review (SLR) aims at identifying, evaluating and interpreting all relevant materials to a specific research question [1]. The procedure of conducting SLR mainly consists of 5 steps: a) identification of Research; b) study selection; c) study quality assessment; d) data extraction; e) data synthesis. SLR typically requires much more efforts compared to traditional reviews [1]. To conduct SLR, researchers have to review a great number of literature collected from various sources, e.g. conference and journal papers, technique reports, advices of authorities, and various other materials. Researchers have to invest considerable efforts to assure the overall quality. For example, in the study selection step, *inclusion/exclusion* decision for each literature must be achieved and disagreements are resolved through discussion among participants. In the data extraction step, participants should carefully review each literature to achieve consistent understanding on its details. Though quick scanning may be helpful in making decisions, it tends to decrease the fairness and objectivity as well as increase the bias of the final results. Therefore, methods that could both reduce the involved effort and guarantee the quality of SLRs are in great need, which motivates the research work of this study.

There are numerous studies focusing on how to effectively and efficiently conduct SLR in software engineering field. Zhang et al. [2] employ a systematic and evidence-based approach to develop and execute optimal search strategies in SLR. However, they have to construct the "standard" manually when conducting SLR each time, which is quite an effort intensive task. Emam et al. [3] tried to apply visualization in support of the conducting of SLR. However, lots of human judgments are also indispensable to implement the SLR processes. This paper reconsiders the problem of using ontology to organize and manage knowledge needed for SLR.

Defined as "an explicit specification of a shared conceptualization." by Gruber [4], ontology has many advantages in dealing with knowledge related works [5]. Ontology-based approaches attract software engineering researchers' attentions in many sub-fields. The whole lifecycle of software engineering could be greatly assisted [6]. Zhang et al. [7] use ontology-based approach to reason security concerns of software. Ontology can be used to organize knowledge of a problem domain [8]. Due to a number of advantages of ontology, knowledge-centered tasks could be assisted by applying ontology techniques. One of these tasks is systematic literature review in software engineering field. We try to model an ontology structure capable of supporting SLR.

In this paper, abstract is an important source for developing our ontology structure. According to Kitchenham et al. [1], current procedures suggest that a review of the title and abstract of a primary study should be sufficient to decide whether or not one study is relevant to a SLR. Budgen [9] has proved that structured abstracts are easier to read and contain more complete information. Kai Petersen [10] also claimed that structured abstracts may be a great help in conducting SLR. Though SLR could not be conducted by reviewing abstract only, abstract provides abundant and significant information that is in support of SLR steps, such as study selection. Therefore, this study chooses abstract to help build ontologies.

The purpose of this study is to find an approach to build general empirical ontology, and specifically address its application on facilitating SLR in the software engineering field. The structure of abstracts and SLR conducting protocol are chosen as an important evidence of the ontology structure. Based on these evidences, *Systematic Literature Review ONTology (SLRONT)* is discussed and built. Then we propose a specification extension of SLRONT on cost estimation systematic review, and build the structure of *COSt estimation ONTology (COSONT)*. Key concepts and knowledge are extracted and added to the ontology as instances to instantiate the ontology. Two case studies illustrate the usefulness of our approach.

The rest of this paper is organized as follows: section 2 presents the details of building SLRONT, extending it to more specific COSONT, instantiating COSONT with extracted knowledge from abstracts and how to use it to support SLR; section 3 describes case studies we conducted; section 4 offers detailed discussion and the limitations; and finally, we draw conclusions in section 5.

## II. METHODOLOGY OF BUILDING ONTOLOGIES

The overview of methodology is shown in Figure 1. First, we establish SLRONT based on typical structure of SLR review protocol and structure of typical domain abstracts, which work as groundwork for more specific ontologies. Then based on SLRONT and specific domain knowledge, we extend SLRONT into more specific ontology, the COSONT. Working as a knowledge base, COSONT could support research query for concrete cost estimation questions. The retrieved concepts and knowledge provides support for SLR procedures.

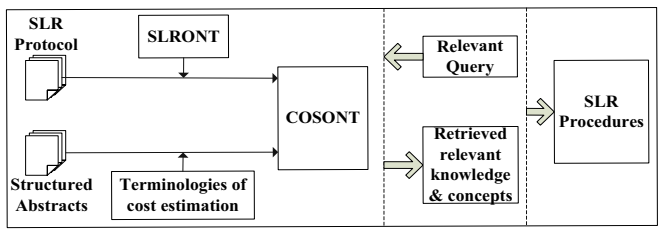

Figure 1. Methodology

### A. Construction of SLRONT

Whether the structure of ontology is suitable enough mainly depends on whether or not it could effectively support knowledge centered tasks in certain domain [5]. In this sub-section, we develop the hierarchy [11] of SLRONT. SLRONT acts as the general base of our approach. All specific domain ontologies should be built by extending it on certain aspects. SLRONT contains two sub-parts, the review protocols and the primary study. The relationship between them is a many-to-many mapping. That is, for a specific systematic literature review protocol, there are many primary studies relating to it and vice versa. Figure 2 shows the top level structure of it.

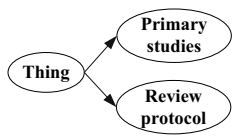

Figure 2. Top level SLRONT

*1) Ontology structure of systematic literature review protocol:* According to Kitchenham et al. [1], SLRs conform to strictly defined procedures. The top level procedure of SLR includes three phrases, planning, conducting and reporting. In order to guarantee the overall quality of the whole procedures, review protocol should be defined according to specific requirements at earliest planning stage. Researchers should build impartial, unbiased protocol based on their expert judgment [1]. They should figure out what the constraints are and why the SLR should be conducted in that way. The protocol is helpful for researchers to understand the research question for the reason that once all these factors are designed, the scope of SLR is also clearly constrained. When added with the review protocol elements, the ontology grows into the following form in Figure 3:

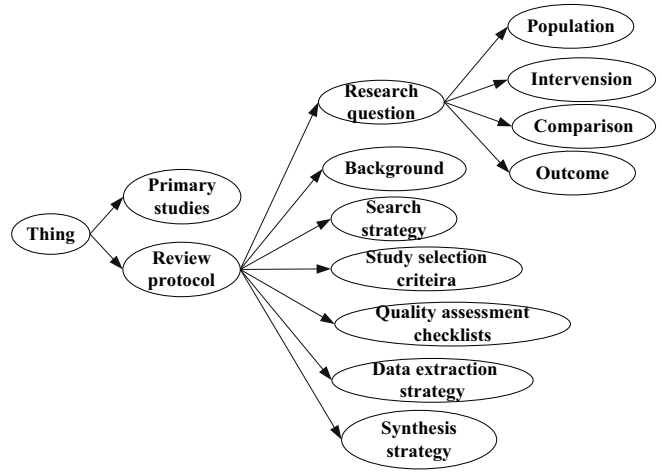

Figure 3. The review protocol class hierarchy

*2) Ontology structure of primary study:* Most primary studies are well written in an consistent modern writing style. Generally they are organized as: title, author & affiliations, abstract, keyword, full-text and references. Almost all of these parts need to be reviewed in SLR in order to find the answer to the research question. In this study, we focus mainly on the abstract part (other parts of literature may be surveyed in our future works), whose quality is assured "inborn" by the authors. Though reviewing pure abstract could not fulfill SLR's requirement of "thoroughly survey every primary study", abstract is still effective enough for narrow down the scope of candidate relevant studies.

Abstract plays a unique role in conducting SLRs. In SLR, the purpose of the study selection step is to identify primary studies that provide direct evidence to the research question [1], which is mainly performed with the help of abstracts in the reality (review the detail of full-text is still necessary if disagreement exists). What's more, data extraction in abstract could assist SLR tasks to some extent because the most crucial information is usually presented in abstract part.

The advocating of writing structured abstracts has been prevailing in recent years. Structured abstract has better clarity and completeness compared to conventional abstracts [12] [13] [9], which may contribute to the conducting of SLR. Besides the quality, structured abstracts support SLR better in that researchers could find more relevant and precise information easily in light of the declaration of structures. For example, if a SLR is designed to survey what is the most popular research approach adopted in a specific domain, the "method" part of structured abstracts should be reviewed mainly.

Though most of the abstracts in published papers' are not written in structured style, they are still organized in a potential logic similar to that of structured abstracts', only the logic is much vaguer. Therefore, researchers could take advantage of this fact and use syntactical analysis techniques to transfer conventional abstracts into structured ones to support further research. This could contribute a lot to the conducting of SLR.

A standard structured abstract includes the following five parts: Background, Object, Method, Result and Conclusion. We develop the following structure to represent primary studies. Details are shown in Figure 4. The "Structured abstracts" class is extended with five sub-classes. The dotted circles represent the parts not surveyed in this study.

structured abstracts into the "Method" part and combine the "Result" part into the "Conclusion" part. Therefore, five subclasses of structured abstracts are briefed into three subclasses: Background, Method and Conclusion.

The main extending work lies in the Method sub-class of structured abstracts. According to the cost estimation SLR of Jorgensen et al. [14], the most common research topic is the introduction and evaluation of methods (with 61% of the papers). The popular methods include Regression, Neural network, Analogy, Expert judgment, Bayesian, etc. Also metrics such as Size and Uncertainty of effort are popular topics. Therefore, we concluded that the most widely concerns of this field are: *a) What methods are used; b) What are the metrics adopted; c) In what context the study is launched.* We extend the SLRONT into COSONT based on the above discussions. Details are shown in Figure 5.

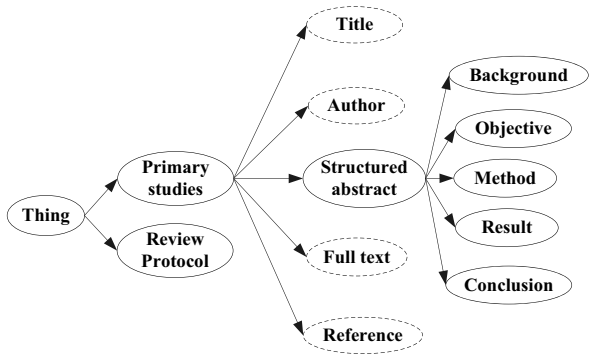

Figure 4. The primary study structure

### B. Extending SLRONT to COSONT

To demonstrate the value of ontology, we extend SLRONT to more specific version, the COSONT. COSONT is a more detailed ontology aiming at supporting cost estimation systematic literature review. Since the quality of unstructured abstracts is not as good as structured ones, we make a simplification while modeling it so that unstructured abstracts could be converted into structured ones with grammatical and syntactical analysis. We combine the "Object" part of

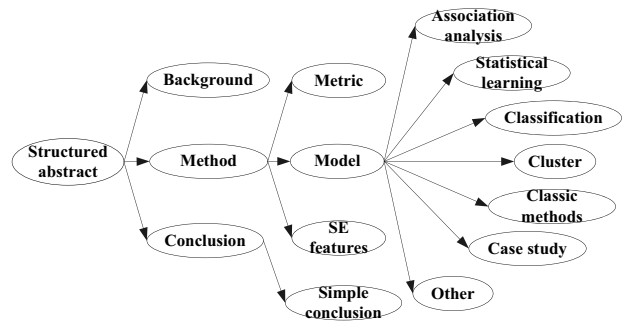

Figure 5. COSONT structure

For the "Method" class, we build three sub-classes: Model, Metric and SE_feature. Due to the limitation of space, only partial subclasses of "Model" class are presented. We also build a "Simple conclusion" class as the subclass of the "Conclusion" class. This is only an empty structure. COSONT needs to be instantiated through analyzing the abstracts of papers.

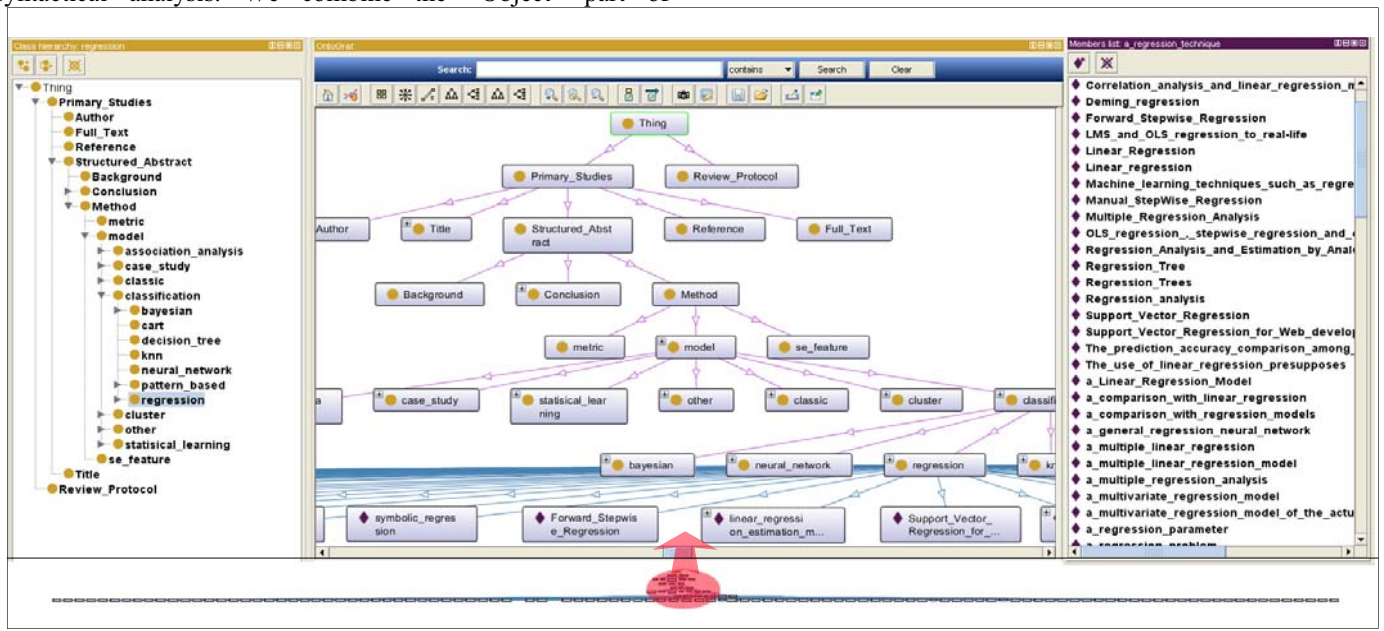

Figure 6. The Instantiated COSONT

First we build a set of grammatical and syntactical rules to segment conventional abstracts into structured abstracts. We use the *Stanford Parser* to parse the sentences of method and conclusion parts to extract knowledge and concepts as well as simple conclusions. The background part is not analyzed for it contains less relevant information towards the main theme of one paper. Extracted knowledge is added to COSONT according to predefined category. More details can be found in our technical report [15]. The instantiated COSONT is shown in Figure 6. Left column is the class hierarchy of COSONT. The bottom part is the entire graphical structure of COSONT expanded on the regression individuals, and its essential part is amplified in the middle zone. Yellow dots denote classes and purple rhombus denote individuals. Right column shows the regression individuals list extracted from structured abstracts.

### C. Automated Query and Retrieval

When COSONT is built up, it works as a knowledge base for further SLR studies. Since key knowledge and concepts relevant to specific cost estimation research question are extracted from abstracts and are inserted into the ontology structure, researchers could use SPAROL queries to retrieve them. Then relevant analysis is conducted on the retrieved results. This will save much more efforts than manually checking each literature and extracting information out of them. Next section will discuss two case studies.

### III. CASE STUDY

Since regression and neural network are two popular estimation approaches, we conduct 2 example scenarios to find which approach is better. Time used, studies identified and final conclusion are evaluating metrics.

1) Research question for scenario 1:

*Find all cost estimation literatures discussing both regression and neural network methods.*

2) Research question for scenario 2:

*Is regression method considered "better" than neural network method?*

Towards the research area of software effort estimation, we use "effort prediction" as the primary search string. Four most popular digital libraries: IEEE Explorer, ACM Digital Library, Springer and Science Direct are searched with this string. As a raw set, we got 645 papers. Then, all parts of abstract are manually examined, with the help of title and keywords. If there are papers difficult to identified, we will go through the full text to make advanced decision. Moreover, experts are invited to supervise the whole process and assure the quality of final results. After carefully selection, 347 papers are selected as the final dataset.

We automatically segment these conventional abstracts into structured abstracts with syntactical analysis, then we instantiate COSONT with extracted knowledge and concepts from method and conclusion parts. Each paper is added to the primary class as an instance. All the noun phrases extracted from the method part are added to classes and subclasses of Model, Metric and SE_feature classes in accordance with the hierarchical categorization. Simple conclusions extracted from conclusion part are added to the Conclusion class as individuals. We invited 4 PhD students to participate in our case study; all of them have background of SLR. At first, we give them proper training to enable them to have an overview of the whole work.

### A. Study 1: supporting study selection

In this step, each student is randomly assigned around 87 papers of the overall 347 papers. The main inclusion/exclusion criterion is that the paper should mention both regression and neural network models. They should go through the details of abstracts and filter out those irrelevant ones based on their personal judgment. They were also asked to record the efforts used. As a comparison, we query COSONT to check the method part of abstracts automatically to find relevant paper. Figure 7 gives the SPARQL query and Table I shows the result.

```
SELECT ?primary_study
FROM <cosont.owl>
WHERE {
    ?selected_studies foaf:model "regression" and
                     foaf:model "neural network"
}
```

Figure 7. SPARQL for finding relevant studies

TABLE I. MANUAL SELECTION OF STUDIES

| Student ID | Num of papers | Paper identified | Time (Person Hour) |
|---|---|---|---|
| 1 | 87 | 6 | 8.5 |
| 2 | 87 | 1 | 7.5 |
| 3 | 87 | 2 | 9 |
| 4 | 86 | 2 | 10 |
| Total | 347 | 11 | 35 |

As shown in table 1, 11 papers are identified at last. Manual selection is rather time consuming. The total time used is 35 *Person Hours*. Using COSONT, we select *the same* 11 studies but time used by COSONT could be ignored. Note that we need time to construct COSONT which should be added to the total time consumed by COSONT approach. While, building COSONT is an automatic procedure and adapts well to scale growing. When there are new abstracts needed to be added, we could simply re-run the procedure and update COSONT, which accounts for the good expandability of ontology based method. Results show that it is rather convenient to use COSONT to accomplish this task. After this study selection step, we give the answer to our SLR question: *There are 11 papers in the dataset discussing both regression and neural network method.*

### B. Study 2: supporting data extraction

In this case study, 4 PhD students need to find whether regression method is better than neural network in cost estimation research area. They should further check abstracts of the 11 identified paper set to judge whether the topic of a paper is about the comparison between two models, and make further judgment about which model is better. Meanwhile, we use COSONT to get simple conclusions extracted from each paper. And we make our decision based on the extracted information. Figure 8 gives the SPARQL query and Table II shows the result. Note that R denotes regression and NN denotes neural network.

```
SELECT ?simple_conclusion
FROM ?selected_studies
```
Figure 8.  SPARQL for finding simple conclusion

TABLE II.  MANUAL DATA EXTRACTION VS. USING COSONT

| Experiment Tester | No. of Abstracts about Comparison | Number of result | | | Which is Better | Time Used (min) |
|---|---|---|---|---|---|---|
| | | NN | R | Almost the same | | |
| Student 1 | 5 | 3 | 0 | 2 | NN | 39 |
| Student 2 | 4 | 3 | 0 | 1 | NN | 31 |
| Student 3 | 4 | 3 | 0 | 1 | NN | 33 |
| Student 4 | 5 | 3 | 0 | 2 | NN | 37 |
| COSONT | 5 | 3 | 0 | 2 | NN | 12 |

As shown in Table II, of the total 11 papers, student 1 believes there are 5 papers about the comparison between two methods. Among them, 3 papers contend neural network is better, no paper said regression is better, and 2 papers said there is no big difference between them. So Student 1 reached the conclusion that neural network is better. The table shows that all of the 5 testers (including COSONT) reach *the same* conclusion. Time used by applying COSONT is much less than each of the 4 PhD students. After this data extraction step, we could give answer to our research question: *Neural network is "better" than regression model in performing cost estimation.*

In these studies, we reached the same conclusion using COSONT compared as sheer manual work. However, in each step of SLR, time used by COSONT is significantly less.

## IV. DISCUSSION

In this study, a general purpose ontology, SLRONT, is built to support systematic literature review. To illustrate its usage and effectiveness, a specialization extension, COSONT is built based on it. There are many benefits of our approach.

First, ontology has multiple advantages in supporting knowledge intensive tasks, such as knowledge expansion, knowledge reuse and machine reasoning. Though extra efforts may be cost in building ontology, the benefits of ontology are profound and profitable in the long run.

Second, ontology structure enables sharing knowledge between different research groups to enhance communication. With the help of ontology, small scale research groups lacking resource could conduct SLRs easily and worry less about paying too many efforts.

Third, our automatic method extracts core knowledge from more relevant parts, such as the method and conclusion part of structured abstracts, and it could quickly filter out irrelevant materials. Knowledge contained in some part of abstract may confuse the "search engine like" searching behavior, for example, irrelevant concepts in background part.

There are also some limitations of our work. First, more detailed ontology should firstly be configured to more specific applications in specific domain. Second, in COSONT, we reduce the five sub-parts of abstracts into three sub-parts. This level of granularity may only works in the context of this study.

Third, we only give an example of facilitating the second and fourth step of systematic literature review. Further studies are needed in proving the usefulness of ontology in supporting other conducting steps. Last but not least, only abstract part is analyzed in our work. The full-text and other parts of primary studies may also necessary to be analyzed.

## V. CONCLUSION

We propose a general purpose SLRONT to facilitate knowledge intensive researches, the systematic literature review. The review protocol of SLR is modeled into the ontology in order to better organize these concepts. Structured abstracts are also modeled to build SLRONT. To illustrate its effectiveness, we specify it into a more detailed structure, the COSONT. Then we use COSONT to support the ponderous work involved in systematic literature review by conducting case studies. The results illustrate the usefulness of our approach and therefore convince us that ontology is a good tool to support intensive efforts required in conducting SLR.